\documentclass[prl, superscriptaddress, showpacs, floatfix, twocolumn]{revtex4-2}
\usepackage[utf8]{inputenc}
\usepackage{graphicx}
\usepackage{hyperref}
\usepackage{amsmath}
\usepackage{amssymb}
\usepackage{makecell}
\usepackage{tabularx}
\usepackage{booktabs}

\hypersetup{
    colorlinks=true,
    linkcolor=blue,
    filecolor=blue,
    urlcolor=blue,
    citecolor=blue,
}

\begin{document}
\title{Intersite Coulomb repulsion driven quadrupole instability and magnetic ordering in the orbital frustrated Ba$_2$MgReO$_6$}
\author{Xuanye Zhang}
\thanks{These authors made equal contributions to this work.}
\affiliation{Wuhan National High Magnetic Field Center $\&$ School of Physics, Huazhong University of Science and Technology, Wuhan 430074, China}
\author{Jinyu Zou}
\thanks{These authors made equal contributions to this work.}
\affiliation{Wuhan National High Magnetic Field Center $\&$ School of Physics, Huazhong University of Science and Technology, Wuhan 430074, China}
\author{Gang Xu}
\email[e-mail address: ]{gangxu@hust.edu.cn}
\affiliation{Wuhan National High Magnetic Field Center $\&$ School of Physics, Huazhong University of Science and Technology, Wuhan 430074, China}
\affiliation{Institute for Quantum Science and Engineering, Huazhong University of Science and Technology, Wuhan, 430074, China}
\affiliation{Wuhan Institute of Quantum Technology, Wuhan, 430074, China}

\begin{abstract}
We develop an unrestricted Hartree-Fock mean-field method including Coulomb repulsion $U$, $V$ and spin-orbital coupling $\lambda$ self-consistently to investigate the mechanism of structural instability and magnetic ordering in Ba$_2$MgReO$_6$. A comprehensive quadrupole phase diagram versus $U$ and $V$ with $\lambda$=0.28eV is calculated. Our results demonstrate that the easy-plane anisotropy and the intersite Coulomb repulsion $V$ must be considered to remove the orbital frustration. The increasing of $V$ to $>$20meV would arrange quadrupole $Q_{x^2-y^2}$ antiparallelly, accompanied with small parallel $Q_{3z^2-r^2}$, and stabilize Ba$_2$MgReO$_6$ into the body-centered tetragonal structure. Such antiparallel $Q_{x^2-y^2}$ provides a new mechanism of Dzyaloshinskii-Moriya interaction, and gives rise to the canted antiferromagnetic (CAF) state along [110] axis. Moreover, sizable octupoles such as $O_{21}^{31}$, $O_{21}^{33}$, $O_{21}^{34}$ and $O_{21}^{36}$ are discovered for the first time in CAF state. Our study not only provides a comprehensive understanding of the experiment results in Ba$_2$MgReO$_6$, but also reveals some commonalities of 5d compounds.
\end{abstract}
\maketitle

\textit{Introduction.}—
5d electrons usually have two remarkable characters. 
One is the dramatically enhanced spin-orbital coupling (SOC) $\lambda$ originated from the huge atomic number Z. 
The other is the spatially extended orbital, which could lead to considerable intersite Coulomb repulsion $V$. 
In the past decades, the interplay between onsite Coulomb repulsion $U$ and SOC with comparable energy and the resulting exotic properties in 5d transition metal (TM) compounds have attracted increasing interest, including the SOC assistant Mott metal-insulator transition~\cite{Kim2008prl,kim2009phase,PhysRevLett.125.097202}, non-collinear magnetic moment~\cite{PhysRevB.86.014428,PRB85045124,PhysRevLett.115.156401,lu2017magnetism}, orbital frustration~\cite{PLRwu,PhysRevLett.89.227203,PRR3033} and high-rank multipoles~\cite{annurev, zhao2016np,Orthogonal2017np,wangprl,PhysRevLett124,qiu2021prl}. 
Among them, the orbital frustrated honeycomb and face-centered cubic (Fcc) lattice magnets have been extensively studied recently, which are reported to host the long-pursuit quantum spin liquid states when the orbital frustration maintains and dominates~\cite{balents2010spin,PhysRevLett105,PRL11310}, or give rise to lattice distortion accompanied with the orbital ordering when orbital frustration is eliminated~\cite{ordereddouble,khomskii2003orbital,PRB98214,cong2019}. 
However, while most efforts have been focused on the interplay between $U$ and $\lambda$ and the resulting properties~\cite{PRB7505,PRB820851,PRL11120,schaffer2016recent,cong2020first,PRB1000,da2022impact,PRR5L012010}, rare works pay attention to the influence of $V$, though it could play a key role to determine the exotic states and properties in 5d TM compounds. 
Especially in the first-principles calculation field, a generic program to deal with $U$, $V$ and $\lambda$ self-consistently is still lacking.

The double perovskite Ba$_2$MgReO$_6$ with 5d$^1$ configuration, adopting the frustrated Fcc lattice, provides an ideal platform to study the interplay between $U$, $\lambda$ and $V$, as well as the resulting exotic properties~\cite{bramnik2003,marjerrison2016cubic,dar2018insight,Zenji2019jps,Zenji2020prr,takayama2021spin,msg2021prb,pressure,svoboda2021prb,dft2021prm,phono2022arx}.
 Ba$_2$MgReO$_6$ undergoes two phase transitions upon cooling, including Fcc to Bct (body-centered tetragonal) phase transition at $T_q$ and canted antiferromagnetic (CAF) phase transition with a small gap at $T_m$ ~\cite{Zenji2019jps,Zenji2020prr,msg2021prb,phono2022arx}.
Further synchrotron X-ray diffraction measurements of high-quality single crystals show the tetragonal distortion is enhanced below $T_m$, which indicates that quadrupoles are coupled with magnetic orders~\cite{Zenji2020prr}.  
Recently, magnetic entropy is obtained by subtracting phonon contribution in heat capacity~\cite{phono2022arx}, which reflects the $N$=2 degeneracy of the ground state multiples, contrary to Ref~\cite{Zenji2019jps}.
These evidences imply that the pure spin model is not enough to describe the magnetic mechanism in Ba$_2$MgReO$_6$.
In fact, strong SOC not only leads to small magnetic dipole ordering, as small as 0.3$\mu_B$ in Ba$_2$MgReO$_6$, but also opens up the possibility of high-rank multipoles~\cite{chen2010prb,wangprl}. 
Such high-rank multipoles are usually entangled with the magnetic dipole ordering together, which makes it hard to detect these hidden orders by conventional experiments~\cite{zhao2016np,Orthogonal2017np}. 
Therefore, a comprehensive understanding of the phase transitions and exotic properties in Ba$_2$MgReO$_6$ remains elusive until now.

In this letter, based on first-principles calculations, we develop an unrestricted Hartree-Fock mean-field method which includes $U$, $\lambda$ and especially $V$ self-consistently to investigate the nature of phase transition and exotic properties in Ba$_2$MgReO$_6$.
Our study demonstrates that, while the easy-plane anisotropy can partially remove the orbital frustration, the intersite Coulomb repulsion V is indispensable to fully remove the frustration and generate the right orbital ordering to meet the experiment.
The calculations figure out that the antiparallel (AP) $Q_{x^2-y^2}$ could minimize the intersite Coulomb repulsion energy mostly.
Hence, small $V$=20meV is enough to stabilize Ba$_2$MgReO$_6$ into the Bct structure accompanied with the AP-$Q_{x^2-y^2}$ and small magnitude of parallel $Q_{3z^2-r^2}$ quadrupole.
Such a quadrupole state with incompletely quenched orbital angular momentum provides a new mechanism of Dzyaloshinskii–Moriya (DM) interaction, resulting in the experimentally observed CAF ground state~\cite{Zenji2019jps,Zenji2020prr}.
Moreover, various magnetic octupoles are discovered in the CAF ground state, where the octupoles have comparable magnitude with magnetic dipoles and also exhibit canted arrangement along [110] direction. 
This complex magnetic structure may be the source of abnormal magnetic entropy and weak magnetic anisotropy in the CAF ground state.

\begin{figure}
  \includegraphics[width=\columnwidth]{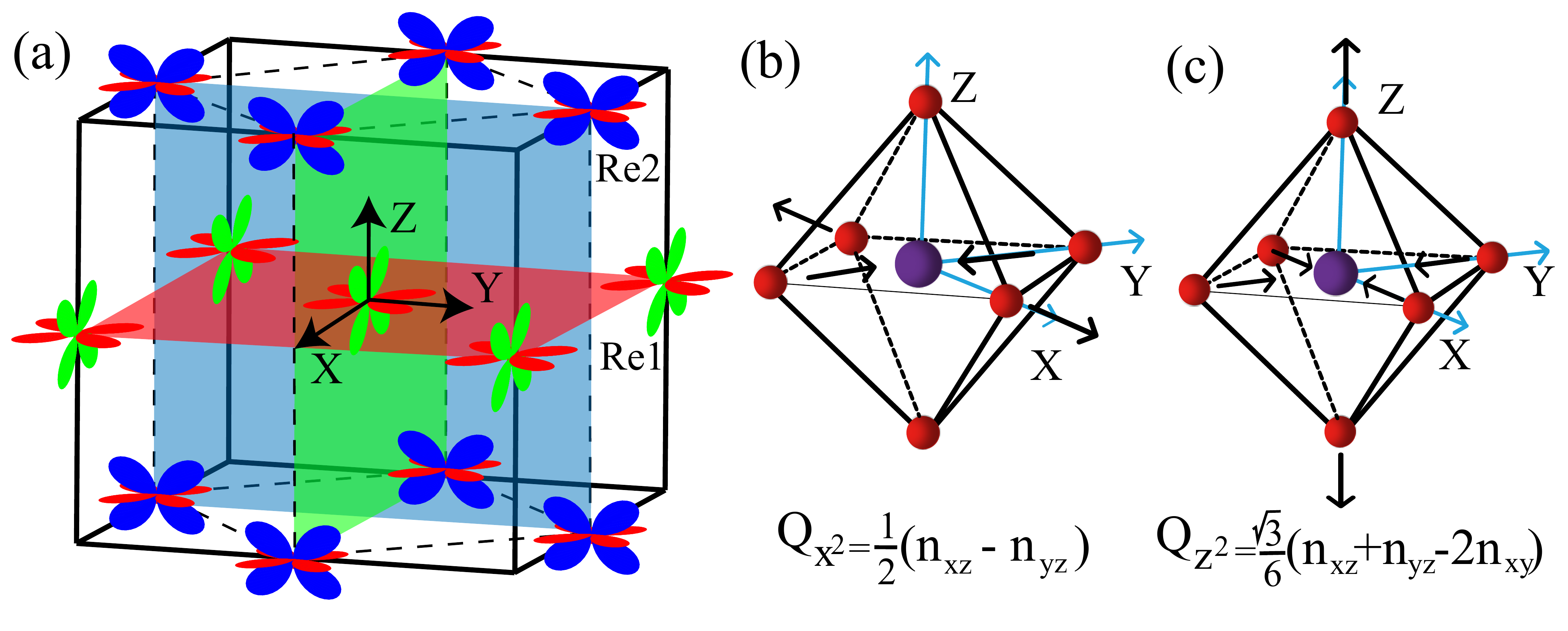}
  \caption{(a) Schematic of Re1 and Re2 atoms in Fcc unit cell (solid line) and Bct structure (dashed lines). XY, XZ and YZ orbital (plane) are indicated by red, green and blue color. (b)(c) The oxygen octahedral distortion modes corresponding to $Q_{x^2}$ and $Q_{z^2}$ orders. 
  }
  \label{f1}
\end{figure}

\textit{Calculation methods.}—
As shown in Fig.~\ref{f1}(a), Ba$_2$MgReO$_6$ adopts a double perovskite structure with $Fm\bar3m$ space group~\cite{takayama2021spin,Zenji2019jps}. The crystal field of the local oxygen octahedron splits d-orbitals of Re atom into higher doubly-degenerate $e_g$ orbitals and lower triply-degenerate $t_{2g}$ orbitals~\cite{maekawa2004physics,egt2g2018prb}. With 5d$^1$ configuration, $t_{2g}$ orbitals are enough to describe low physics of Ba$_2$MgReO$_6$. The one electron occupation usually leads to spontaneous symmetry breaking with orbital ordering~\cite{kugel1973crystal,kugel1982spu,PhysRevLett82,PhysRevLett98,PhysRevB83165,PhysRevB9722}, which can be expanded by the quadrupole orders $Q_{x^2-y^2}$ (abbr. $Q_{x^2}$) and $Q_{3z^2-r^2}$ (abbr. $Q_{z^2}$)
\begin{equation}\label{eq:Q}
\begin{aligned}
Q_{x^2}&=\frac{1}{2}(n_{xz}-n_{yz})\\
Q_{z^2}&=\frac{\sqrt{3}}{6}(n_{xz}+n_{yz}-2n_{xy}),
\end{aligned}
\end{equation}
where $n_\alpha$($\alpha$ =xy, xz, yz) is electron occupation. The $Q_{x^2}$ breaks $O_h$ to $D_{2h}$ and $Q_{z^2}$ breaks to $D_{4h}$, corresponding to the octahedral distortion shown in Fig.~\ref{f1}(b) and Fig.~\ref{f1}(c) respectively.

\begin{figure}
  \includegraphics[width=\columnwidth]{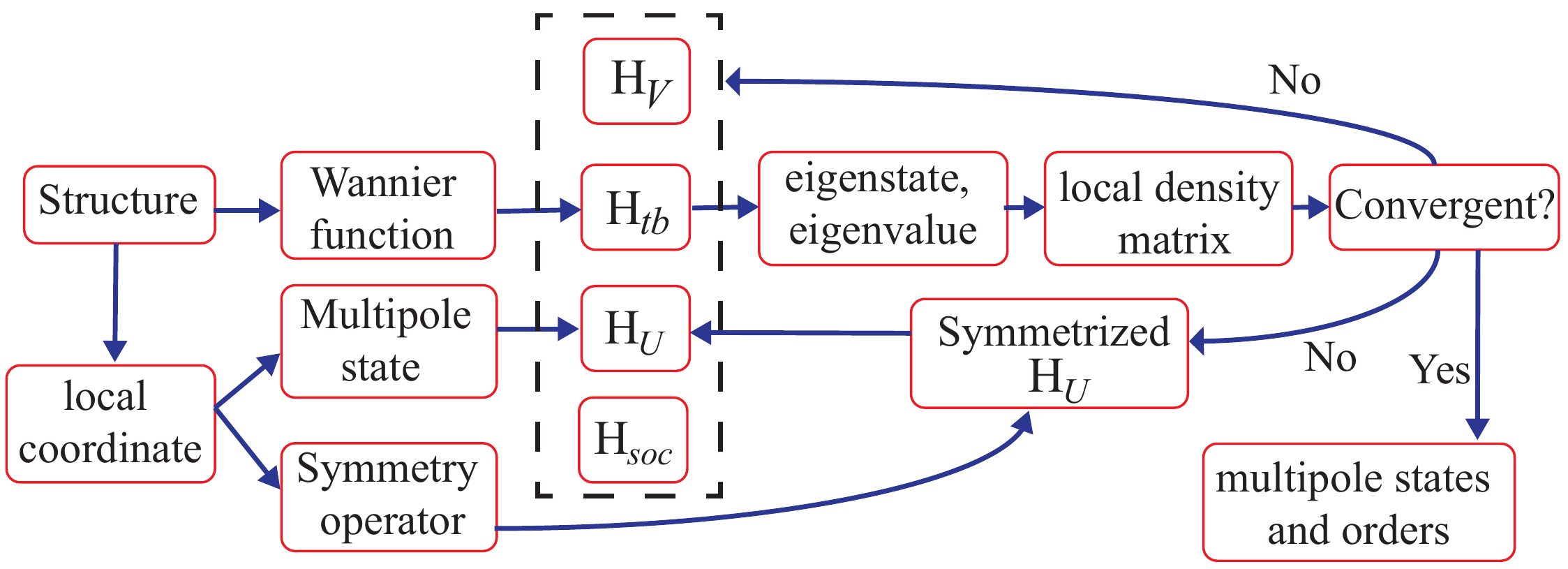}
  \caption{ The flow chart to calculate the multipole orders based on the first-principles calculations. }
  \label{f2}
\end{figure}

In order to comprehensively investigate the effects of intersite Coulomb repulsion $V$ on the orbital ordering and physical properties in Ba$_2$MgReO$_6$, we develope a self-consistent calculation package that includes $V$ to solve a complete Hamiltonian $H=H_{tb}+H_{SOC}+H_U+H_V$, as shown in Fig.~\ref{f2}. Starting from the Fcc structure with lattice constant $a$=8.0802$\textrm{\AA}$~\cite{Zenji2020prr}, the tight-binding Hamiltonian $H_{tb}$ is obtained by constructing the maximum localized wannier function of $t_{2g}$ using the Vienna ab initio simulation ~\cite{PhysRevB5411,KRESSE199615} and WANNIER90 package ~\cite{MOSTOFI20142309,wannier90,maximally}. The spin orbital coupling $H_{SOC}$ adopts the form of $\lambda \boldsymbol{S} \cdot \boldsymbol{L}_{t_{2g}}$, with the spin and effective orbital angular momentum $S=\frac{1}{2}$ and $L_{t_{2g}}=1$. $\lambda$=0.28eV is determined by the non-magnetic (NM) band structures fitting.

The onsite interaction $H_U$ adopts the form of Kanamori Hamiltonian
\begin{equation}\label{eq:HU}
\begin{aligned}
H_U=(U-3J)\frac{\hat{N}(\hat{N}-1)}{2}-2J\boldsymbol{S}^2-\frac{J}{2}\boldsymbol{L}^2+\frac{5}{2}J\hat{N},
\end{aligned}
\end{equation}
with the occupation number $\hat{N}=\sum_{\alpha\sigma}\hat{n}_{\alpha\sigma}$ and Hund’s coupling $J/U=0.1$. Applying the mean field approximation, the correlation interaction can be solved directly by the unrestricted Hartree-Fock method:
\begin{equation}\label{eq:HU_MF}
\begin{aligned}
c^\dagger_\alpha c^\dagger_\beta c_\delta c_\gamma &\approx \rho_{\beta\delta} c^\dagger_\alpha c_\gamma + \rho_{\alpha\gamma} c^\dagger_\beta c_\delta -\rho_{\beta\delta}\rho_{\alpha\gamma}\\
                                                  &- \rho_{\beta\gamma} c^\dagger_\alpha c_\delta - \rho_{\alpha\delta} c^\dagger_\beta c_\gamma +\rho_{\beta\gamma}\rho_{\alpha\delta},
\end{aligned}
\end{equation}
where $\rho_{\alpha\beta} = \langle c^\dagger_\alpha c_\beta \rangle$ is the local density matrix~\cite{Zhuang1997}. 

The intersite repulsion of $t_{2g}$ is $H_V=\sum_{\langle ij \rangle} V_{ij}^{\alpha\beta} n_{i,\alpha} n_{j,\beta}$. According to electric quadrupole approximation~\cite{chen2010prb}, $V^{\alpha\beta}_{ij}$ can be reduced to a single variable $V$. Considering the d$^1$ configuration and ignoring the constant, it becomes 
\begin{equation}\label{eq:HV}
\begin{aligned}
H_V=V\sum_\alpha \sum_{\langle ij \rangle \in \alpha} [\frac{4}{3}(n_{i,\beta}-n_{i,\gamma})(n_{j,\gamma}-n_{j,\beta})+\frac{4}{9}n_{i,\alpha}n_{j,\alpha}],
\end{aligned}
\end{equation}
in which $\alpha\ne\beta\ne\gamma$ and $\langle ij \rangle \in \alpha$ denote that the bond between $i$ and $j$ is in the plane of $\alpha$ orbital. With the mean field approximation, $H_V$ is solved as
\begin{equation}\label{eq:HV_MF}
\begin{aligned}
H_V=V\sum_\alpha \sum_{\langle ij \rangle \in \alpha}[&\frac{4}{3}(\overline{n}_{j,\gamma}-\overline{n}_{j,\beta})(n_{i,\beta}-\frac{\overline{n}_{i,\beta}}{2}-n_{i,\gamma}+\frac{\overline{n}_{i,\gamma}}{2})\\
                                                   +&\frac{4}{9}(\overline{n}_{j,\alpha} n_{i,\alpha}-\frac{\overline{n}_{j,\alpha}\overline{n}_{i,\alpha}}{2})+(i \leftrightarrow j)],
\end{aligned}
\end{equation}
where $\overline{n}=\langle n \rangle$ can be obtained directly from the local density matrix.

To release the orbital frustration of cubic symmetry, our calculation is performed in the Bct structure as illustrated by the dashed line in Fig.~\ref{f1}(a). This superstructure has the same lattice as the cubic but with the vector Q=2$\pi$(001), which has been widely reported in double perovskite materials~\cite{stitzer2002crystal,yamamura2006structural,PhysRevLett9901,morrow2016spin,PhysRevB.101.220412}. The flow of the calculation program has been plotted in Fig.~\ref{f2}. For the reason that $H_V$ is obtained by local density matrix, the Hamiltonian in the first loop only contains $H_{tb}$, $H_{SOC}$ and $H_U$ which is set according to multipole states and symmetry requirments. Thus, the Hamiltonian is diagonalized to calculate the local density matrix and further abtain the order parameters as well as the expression of $H_V$ for the next loop. The energy accuracy used for our calculation is set as $10^{-6}$eV.

\begin{figure}
  \includegraphics[width=\columnwidth]{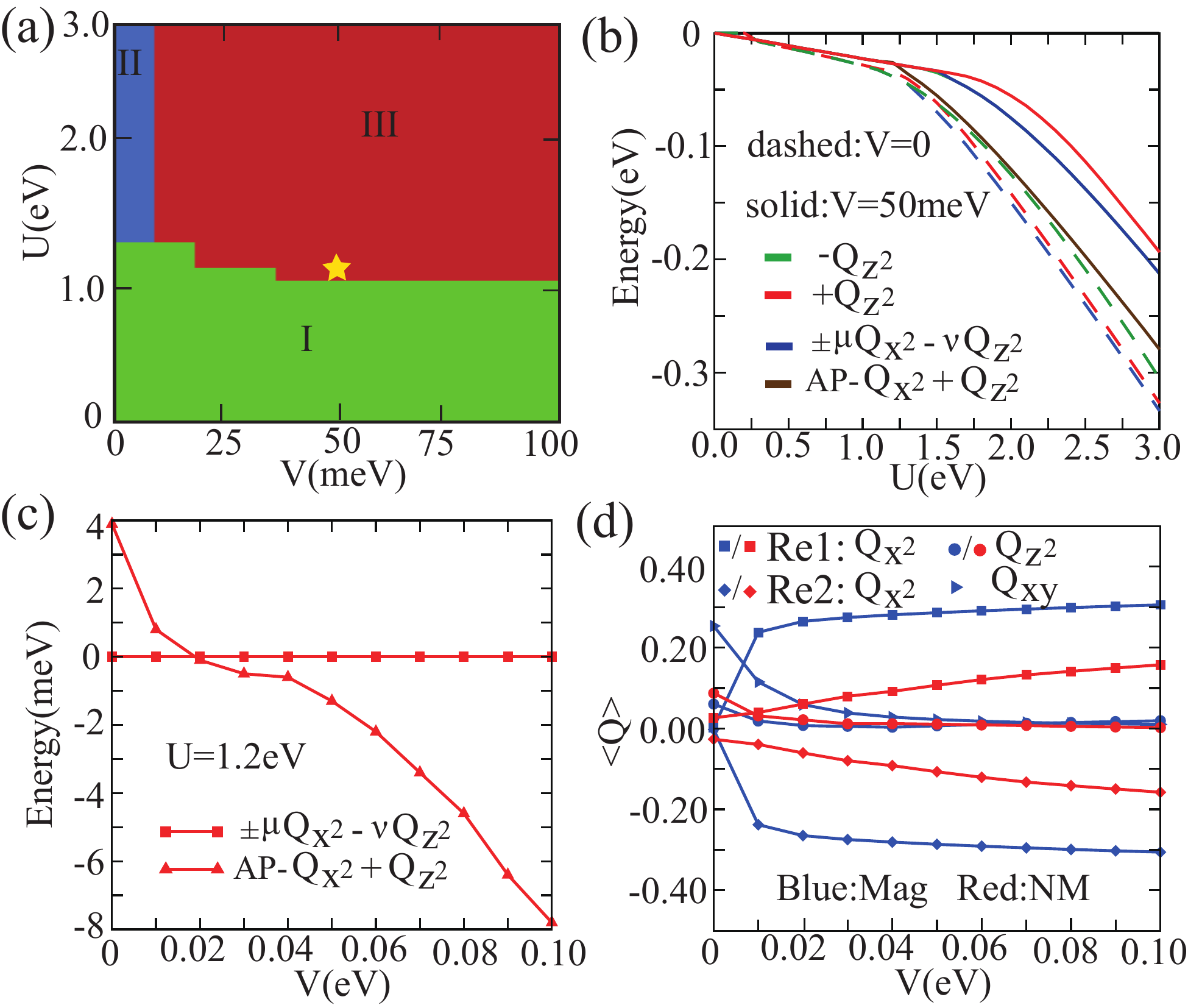}
  \caption{(a) The quadrupole phase diagram versus U and V. The SOC strength is $\lambda$=0.28eV, and Hound' coupling is taken as $J/U=0.1$. Region I, II, and III represent cubic, $\pm\mu Q_{x^2}-\nu Q_{z^2}$, and (AP-$Q_{x^2})+Q_{z^2}$  states, respectively. Ba$_2$MgReO$_6$ is located at the position marked by the yellow star. (b) The total energy of ordering states depending on $U$ with $V$=0 (dashed lines) and $V$=50meV (solid lines). (c) The energy difference between $\pm\mu Q_{x^2}-\nu Q_{z^2}$ and (AP-$Q_{x^2})+Q_{z^2}$ states with the increasing of $V$. 
(d) Evolution of quadruple orders in (AP-$Q_{x^2})+Q_{z^2}$ state with V increasing. Red and blue represent the NM and CAF magnetic states, respectively. $U$ is set as 1.2eV in (c) and (d).}
  \label{f3}
\end{figure}

\textit{Results of quadrupole.}—
We first perform the NM calculation to study the quadrupole orders. Fig.~\ref{f3}(a) displays the phase diagram as function of $U$ and $V$. The order is absent in small $U$ regime (phase I), while emerges at large $U$ (phase II with quadrupole orders $\pm\mu Q_{x^2}-\nu Q_{z^2}$ and phase III with quadrupole order (AP-$Q_{x^2}$)+$Q_{z^2}$, see details below). This result agrees with the reports that the onsite Coulomb repulsion $U$ is the driving force of the orbital ordering in many materials~\cite{PhysRevLett93,PhysRevLett9706,PhysRevB.74.195115,PhysRevLett1022,PhysRevB910451,PhysRevB100041,PRB931551}.

Dashed lines in Fig.~\ref{f3}(b) compare the energy of quadrupole orders as function of $U$ with $V=0$, which find that doubly-degenerat orders $+\mu Q_{x^2}-\nu Q_{z^2}$ and $-\mu Q_{x^2}-\nu Q_{z^2}$($\mu\neq\frac{\sqrt{3}}{2}$ and $\nu\neq\frac{1}{2}$) become the most stable states when $U>$ 1.2eV (phase II). Such orders reflect that the electron in Ba$_2$MgReO$_6$ tends to occupy $d_{xy}$ orbital, which implies the easy-plane anisotropy as also mentioned in previous work~\cite{chen2010prb}.
The orbital frustration is partially broken in phase II, which is insufficient to interpret the non-degenerate quadrupole order in experiments \cite{Zenji2020prr}.

However, the introduction of $V$ can drive the material into phase III, and give rise to the experimental reported antiparellel quadrupole order \cite{Zenji2020prr} as shown in Fig.~\ref{f3}(c), which demonstrates that the antiparallel $Q_{x^2}$ combined with a small parallel $Q_{z^2}$ order (abbr. (AP-$Q_{x^2}$)+$Q_{z^2}$) repalces $\pm\mu Q_{x^2}-\nu Q_{z^2}$ to become the ground state when $V>20$meV. Such order would lead to the occupation as sketched in Fig.~\ref{f1}(a), which shows that electron on Re1/Re2 atom prefers more $d_{xz}$/$d_{yz}$ orbital and less $d_{xy}$, so that the energy caused by $V$ would be efficiently reduced through keeping the eletron on Re1 and Re2 away from each other. Such mechanism is confirmed by the increment of the alternative occupation of $d_{xz}$ and $d_{yz}$ orbitals on Re1 and Re2 and the reduction of the $d_{xy}$ with the increasing of $V$, giving rise to the increment/reduction of AP-$Q_{x^2}$ order/$Q_{z^2}$ order, as demonstrated by the red lines in Fig.~\ref{f3}(d). 

All these results demonstrate that the intersite Coulomb repulsion $V$ is the key factor for 5d compounds to obtain the right order, and such order is stable against the increasing of $U$ as shown by the solid lines in Fig.~\ref{f3}(b), which present the energy evolution of different orders and clearly demonstrate that (AP-$Q_{x^2}$)+$Q_{z^2}$ order is stabilized when $U>$ 1eV and its energy advantage grows with the increasing of $U$.


\begin{figure}
  \includegraphics[width=\columnwidth]{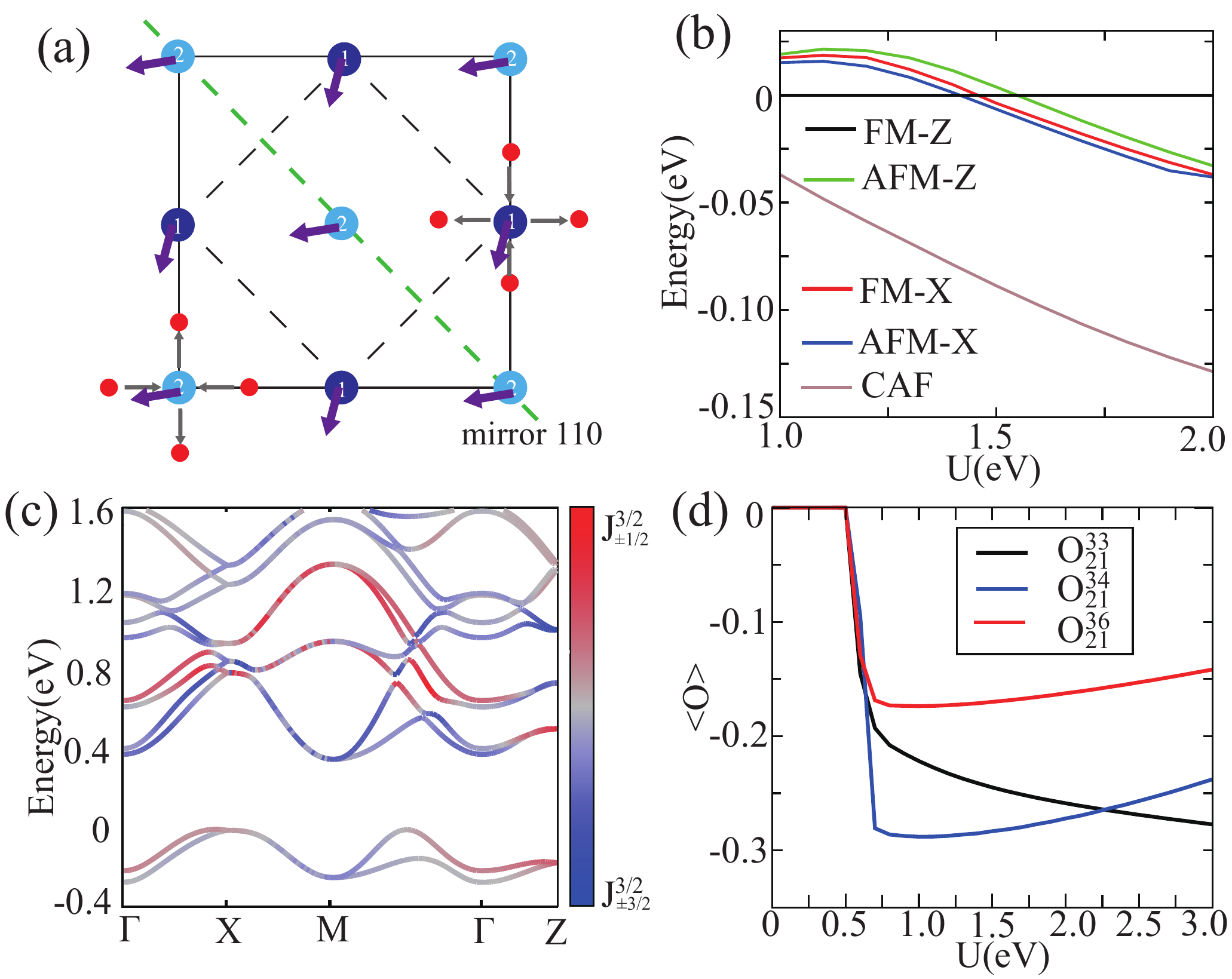}
  \caption{(a) The schematic illustration of the glide mirror symmetry $\{M_{110} |(0,a/2,a/2)\}$ and the CAF magnetic configuration. From the top view of the lattice, the Re1 atoms and Re2 atoms are labeled as 1 and 2. The red small balls represent the oxygen atoms, with the thin arrows represent the distortion of the octahedral corresponding to the AP-$Q_{x^2}$ order. The green dashed line illustrate the plane of the mirror reflection $M_{110}$. The thick violet arrows illustrate the magnetic momentum on Re atoms respecting $\{M_{110} |(0,a/2,a/2)\}$ symmetry.
 (b) Total energy of magnetic states relative to FM-Z with the increasing of $U$. (c)  Band structures of CAF ground state at $U$=1.1eV. (d) The magnitude of Re1's octupoles in CAF ground state depending on $U$, inculding $O^{33}_{21}$, $O^{34}_{21}$ and $O^{36}_{21}$. All the calculations are performed with $V$=50meV.}
  \label{f4}
\end{figure}

\textit{Results of magnetic order.}—
In the following, we perform the calculation with the broken of time reversal symmtry to clearify that (AP-$Q_{x^2}$)+$Q_{z^2}$ order would give rise to the experimental reported CAF configuration. 
From the AP-$Q_{x^2}$ order and the expression $Q_{x^2}=\frac{1}{2}(L_x^2-L_y^2)$ with $L_x=id_{xz}^\dagger d_{xy}+h.c.$ and $L_y=id_{xy}^\dagger d_{yz}+h.c.$~\cite{wangprl}, one can notice that the $L_x$ and $L_y$ on two Re atoms are exchanged ($L_{x,y}$ of Re1$\leftrightarrow$$L_{y,x}$ of Re2) and connected by a glide mirror symmetry $\{M_{110} |(0,a/2,a/2)\}$ as demonstrated in Fig.~\ref{f4}(a). Thus, the angular momentums form a canted angle, which provides a new mechanism of DM interaction $\sum_{<i,j>}D(\boldsymbol{S}_i\times \boldsymbol{S}_j)$ in spin space\cite{PhysRevLett9505,PhysRev.120.91,DZYALOSHINSKY1958241}, as SOC considered. Such DM interaction is an intrinsic property of orbital frustration and would induce the CAF along [110] easy-axis, as confirmed by the self-consistent magnetic calculations in Fig.~\ref{f4}(b) which demonstrate that the ground CAF state is about 75 meV lower  than the other magnetic configurations.

The results of CAF state reproduce the experimental observed dipole and quadrupole orders well. The ratio between $L_x$+2$S_x$ (0.296$\mu_B$) and $L_y$+2$S_y$ (0.128$\mu_B$) of Re1 is approximately 2.3, leading to 0.322$\mu_B$ magnetic moment with canted angle $\phi$$\sim$22$^{\circ}$ along [110] direction, which agree with 0.3$\mu_B$ and $\phi$$\sim$40$^{\circ}$ in experiments \cite{Zenji2020prr}. 
Moreover, CAF state displays obviously enhanced $Q_{x^2}$ order comparing to that of NM state as compared in Fig.~\ref{f3}(d), which coincides well with the experiment report about quadrupole increment below $T_m$ \cite{Zenji2020prr}.

In Fig.~\ref{f4}(c) with $U$=1.1eV, $\lambda$=0.28eV and $V$=50meV, the band structures of CAF ground state gives rise to a Mott insulating state and the band gap about 0.33eV, which agree with the experimental observation (0.17eV)~\cite{Zenji2019jps}. The projection in Fig.~\ref{f4}(c) also demonstrates that the occupied bands have more $|J_{eff}=3/2,m=\pm1/2>$ (red) components than $|J_{eff}=3/2,m=\pm3/2>$ (blue), which agree with the (AP-$Q_{x^2}$)+$Q_{z^2}$ order. We use these fitting parameters to determine the position of Ba$_2$MgReO$_6$ in the quadrupole phase diagram as shown by the yellow star in Fig.~\ref{f3}(a), which is close to the phase boundary between phase III and phase I. This result gives a possible explanation of the  quadrupole vanishing and magnetic phase transition under pressure in Ba$_2$MgReO$_6$\cite{pressure}.

Furthermore, our calculations uncover that the CAF state of Ba$_2$MgReO$_6$ also has considerable octupoles including $O_{21}^{31}$, $O_{21}^{33}$, $O_{21}^{34}$ and $O_{21}^{36}$. We plot the evolution of octupoles on Re1 with the increasing of $U$ in Fig.~\ref{f4}(d), which illustrates the emergence and stabilization of the octupoles when $U>0.5$eV. According to the symmetry of these octupoles \cite{wangprl}, the $\{M_{110} |(0,a/2,a/2)\}$ operation of the AP-$Q_{x^2}$ state enforces the relationship of the octupoles on the two Re atoms as
$O^{31}_{21}$(Re2)=$O^{34}_{21}$(Re1) and $O^{33}_{21}$(Re1)=$-O^{36}_{21}$(Re2), as listed in Table~\ref{table:table1}. Therefore, the AP-$Q_{x^2}$ makes octupoles also have a canted ordering. Since the magnitude of octupoles is as large as dipoles, they should have significant influence on the responses to the external field. It would be interesting and important to detect and control such canted octupoles in the future study of Ba$_2$MgReO$_6$. 

\begin{table}[!t]
 \renewcommand*{\arraystretch}{1.5}
 \centering
\caption{The magnitude of magnetic octupoles at $\lambda$=0.28eV, $U$=1.1eV and $V$=50meV in CAF ground state.}
 \begin{tabular*}{\columnwidth}{c @{\extracolsep{\fill}}ccccc}
\toprule
\hline
    &  $O^{31}_{21}$ &  $O^{33}_{21}$  &  $O^{34}_{21}$ & $O^{36}_{21}$  \\
\hline
         Re1&  0.0051 &  -0.2276  &  -0.2880 & -0.1735   \\
\hline
        Re2& -0.2880 &  0.1735  &  0.0051 & 0.2276 \\
\hline
\bottomrule
\end{tabular*}
 \label{table:table1}
\end{table}

\textit{Conclusion.}—
In summary, we consider the intersite Coulomb interaction $V$ in the self-consistent calculations for the first time to investigate the phase transition mechanism and ground state properties in Ba$_2$MgReO$_6$. The orbital frustration of the cubic structure can be partially lifted by the anisotropy and further fully lifted by $V$. Our calculations demonstrate that $\pm\mu Q_{x^2}-\nu Q_{z^2}$ states are 8meV lower than the $Q_{z^2}$ state in the Bct structure, suggesting the native easy-plane anisotropy favors $d_{xy}$ occupation. The effect of $V$ is opposite to that of the easy-plane anisotropy, which will suppress the $d_{xy}$ occupation and lead to the alternative majority of $d_{yz}$ and $d_{xz}$ orbitals along [001] direction. As a result, the system is stabilized in the (AP-$Q_{x^2})+Q_{z^2}$ quadrupole state when $V$ is larger than 20meV. We uncover that this AP-$Q_{x^2}$ ordering can induce DM interaction between Re1 and Re2 and finally results in the CAF ground state along [110] direction, consistent with the experiments very well. Besides, our calculations discover the additional octupoles $O_{21}^{31}$, $O_{21}^{33}$, $O_{21}^{34}$ and $O_{21}^{36}$ in the CAF state, which also present canted angle with considerable magnitude. 
These results build up a profound understanding between the interactions and the structural instability, magnetic properties, as well as octupoles in Ba$_2$MgReO$_6$. The program developed in this work provides a generic powerful first-principles tool for accurate investigation of other 5d TM compounds.

\textit{Acknowledgments} --- The authors thank Xi Dai, Yilin Wang, Zhida Song, Jianzhou Zhao, Wenxuan Qiu and Aiyun Luo for valuable discussion. This work was supported by the National Key Research and Development Program of China (2018YFA0307000), and the National Natural Science Foundation of China (12274154).

%

\end{document}